\documentclass[sigconf]{acmart}

\usepackage{booktabs} 
\usepackage{gensymb} 
\usepackage{subcaption}
\usepackage{wrapfig,lipsum,booktabs}
\DeclareMathOperator{\sign}{sign}

\setcitestyle{square} 

\usepackage[ruled]{algorithm2e} 

\SetAlFnt{\small}
\SetAlCapFnt{\small}
\SetAlCapNameFnt{\small}
\SetAlCapHSkip{0pt}

\copyrightyear{2021} 
\acmYear{2021} 
\setcopyright{acmlicensed}
\acmConference[SCA '21]{The ACM SIGGRAPH / Eurographics Symposium on Computer Animation}{September 6--9, 2021}{Virtual Event, USA}
\acmBooktitle{The ACM SIGGRAPH / Eurographics Symposium on Computer Animation (SCA '21), September 6--9, 2021, Virtual Event, USA}
\acmPrice{15.00}
\acmDOI{10.1145/3475946.3480950}
\acmISBN{978-1-4503-8676-0/21/09}

\begin{document}
\title{DASH: Modularized Human Manipulation Simulation with Vision and Language for Embodied AI}


\author{Yifeng Jiang}
\affiliation{%
 \institution{Stanford University}
 \country{United States of America}
}
\email{yifengj@stanford.edu}

\author{Michelle Guo}
\affiliation{%
 \institution{Stanford University}
 \country{United States of America}
}
\email{mguo95@stanford.edu}

\author{Jiangshan Li}
\affiliation{%
 \institution{Stanford University}
 \country{United States of America}
}
\email{jiangsli@stanford.edu}

\author{Ioannis Exarchos}
\affiliation{%
 \institution{Stanford University}
 \country{United States of America}
}
\email{exarchos@stanford.edu}

\author{Jiajun Wu}
\affiliation{%
 \institution{Stanford University}
 \country{United States of America}
}
\email{jiajunwu@cs.stanford.edu}

\author{C. Karen Liu}
\affiliation{%
 \institution{Stanford University}
 \country{United States of America}
}
\email{karenliu@cs.stanford.edu}

\renewcommand\shortauthors{Jiang, Guo, Li, Exarchos, Wu, and Liu}




\definecolor{darkblue}{rgb}{0.0, 0.0, 0.53}
\newcommand{\old}[1]{\textcolor{red}{\textbf {\st{#1}}}}
\newcommand{\new}[1]{\textcolor{red}{#1}}
\newcommand{\note}[1]{\cmt{Note: #1}}
\newcommand{\karen}[1]{\textcolor{red}{{[Karen: #1]}}}
\newcommand{\mguo}[1]{\textcolor{orange}{{[Michelle: #1]}}}
\newcommand{\original}[1]{\textcolor{blue}{{[Original: #1]}}}
\newcommand{\jw}[1]{\textcolor{blue}{{[Jiajun: #1]}}}
\newcommand{\yannis}[1]{\textcolor{cyan}{{[Yannis: #1]}}}
\newcommand{\yifeng}[1]{\textcolor{darkblue}{{[Yifeng: #1]}}}
\newcommand{\newtext}[1]{#1}
\newcommand{\eqnref}[1]{Equation~(\ref{eqn:#1})}

\newcommand{\modelfull}{dynamic and autonomous simulated human\xspace}
\newcommand{\model}{DASH\xspace}
\long\def\ignorethis#1{}

\newcommand{\sect}[1]{Section~\ref{#1}}
\newcommand{\myparagraph}[1]{\vspace{5pt}\noindent\textbf{#1}}
\makeatletter
\DeclareRobustCommand\onedot{\futurelet\@let@token\@onedot}
\def\@onedot{\ifx\@let@token.\else.\null\fi\xspace}

\def\iid{i.i.d\onedot}
\def\eg{e.g\onedot} \def\Eg{E.g\onedot}
\def\ie{i.e\onedot} \def\Ie{I.e\onedot}
\def\cf{\emph{c.f}\onedot} \def\Cf{\emph{C.f}\onedot}
\def\etc{etc\onedot} \def\vs{vs\onedot}
\def\wrt{w.r.t\onedot} \def\dof{d.o.f\onedot}
\def\etal{et al\onedot}
\makeatother

\newcommand{\figtodo}[1]{\framebox[0.8\columnwidth]{\rule{0pt}{1in}#1}}
\newcommand{\figref}[1]{Figure~\ref{fig:#1}}
\newcommand{\secref}[1]{Section~\ref{sec:#1}}

\newcommand{\vc}[1]{\ensuremath{\boldsymbol{#1}}}
\newcommand{\pd}[2]{\ensuremath{\frac{\partial{#1}}{\partial{#2}}}}
\newcommand{\pdd}[3]{\ensuremath{\frac{\partial^2{#1}}{\partial{#2}\,\partial{#3}}}}

\newcommand{\vEndEff}{\ensuremath{\vc{d}}}
\newcommand{\vRelMove}{\ensuremath{\vc{r}}}
\newcommand{\sSet}{\ensuremath{S}}

\newcommand{\vControl}{\ensuremath{\vc{u}}}
\newcommand{\vPoint}{\ensuremath{\vc{p}}}
\newcommand{\sSpringCoef}{{\ensuremath{k_{s}}}}
\newcommand{\sDamperCoef}{{\ensuremath{k_{d}}}}
\newcommand{\vHandle}{\ensuremath{\vc{h}}}
\newcommand{\vForce}{\ensuremath{\vc{f}}}

\newcommand{\mTransChain}{\ensuremath{\vc{W}}}
\newcommand{\mRotateTrans}{\ensuremath{\vc{R}}}
\newcommand{\sJoint}{\ensuremath{q}}
\newcommand{\vJoint}{\ensuremath{\vc{q}}}
\newcommand{\mJoint}{\ensuremath{\vc{Q}}}
\newcommand{\mMass}{\ensuremath{\vc{M}}}
\newcommand{\sMass}{\ensuremath{{m}}}
\newcommand{\vGravity}{\ensuremath{\vc{g}}}
\newcommand{\vConstr}{\ensuremath{\vc{C}}}
\newcommand{\sConstr}{\ensuremath{C}}
\newcommand{\vCOM}{\ensuremath{\vc{x}}}
\newcommand{\sGeneralForce}[1]{\ensuremath{Q_{#1}}}
\newcommand{\vStateVar}{\ensuremath{\vc{y}}}
\newcommand{\vControlVar}{\ensuremath{\vc{u}}}
\newcommand{\argmax}{\operatornamewithlimits{argmax}}
\newcommand{\argmin}{\operatornamewithlimits{argmin}}

\newcommand{\tr}[1]{\ensuremath{\mathrm{tr}\left(#1\right)}}

%
%

\renewcommand{\choose}[2]{\ensuremath{\left(\begin{array}{c} #1 \\ #2 \end{array} \right )}}

\newcommand{\gauss}[3]{\ensuremath{\mathcal{N}(#1 | #2 ; #3)}}

\newcommand{\pctab}{\hspace{0.2in}}
\newenvironment{pseudocode} {\begin{center} \begin{minipage}{\textwidth}
                             \normalsize \vspace{-2\baselineskip} \begin{tabbing}
                             \pctab \= \pctab \= \pctab \= \pctab \=
                             \pctab \= \pctab \= \pctab \= \pctab \= \\}
                            {\end{tabbing} \vspace{-2\baselineskip}
                             \end{minipage} \end{center}}
\newenvironment{items}      {\begin{list}{$\bullet$}
                              {\setlength{\partopsep}{\parskip}
                                \setlength{\parsep}{\parskip}
                                \setlength{\topsep}{0pt}
                                \setlength{\itemsep}{0pt}
                                \settowidth{\labelwidth}{$\bullet$}
                                \setlength{\labelsep}{1ex}
                                \setlength{\leftmargin}{\labelwidth}
                                \addtolength{\leftmargin}{\labelsep}
                                }
                              }
                            {\end{list}}
\newcommand{\newfun}[3]{\noindent\vspace{0pt}\fbox{\begin{minipage}{3.3truein}\vspace{#1}~ {#3}~\vspace{12pt}\end{minipage}}\vspace{#2}}

\newcommand{\norm}[1]{\left\lVert#1\right\rVert}

\newcommand{\key}{\textbf}
\newcommand{\fun}{\textsc}



\begin{abstract}

Creating virtual humans with embodied, human-like perceptual and actuation constraints has the promise to provide an integrated simulation platform for many scientific and engineering applications. We present Dynamic and Autonomous Simulated Human (DASH), an embodied virtual human that, given natural language commands, performs grasp-and-stack tasks in a physically-simulated cluttered environment solely using its own visual perception, proprioception, and touch, without requiring human motion data. By factoring the \model system into a vision module, a language module, and manipulation modules of two skill categories, we can mix and match analytical and machine learning techniques for different modules so that \model is able to not only perform randomly arranged tasks with a high success rate, but also do so under anthropomorphic constraints and with fluid and diverse motions. The modular design also favors analysis and extensibility to more complex manipulation skills.

\end{abstract}


%
%


\maketitle

\section{Introduction}

\begin{figure}[t]
\centering
\includegraphics[width=1.0\linewidth]{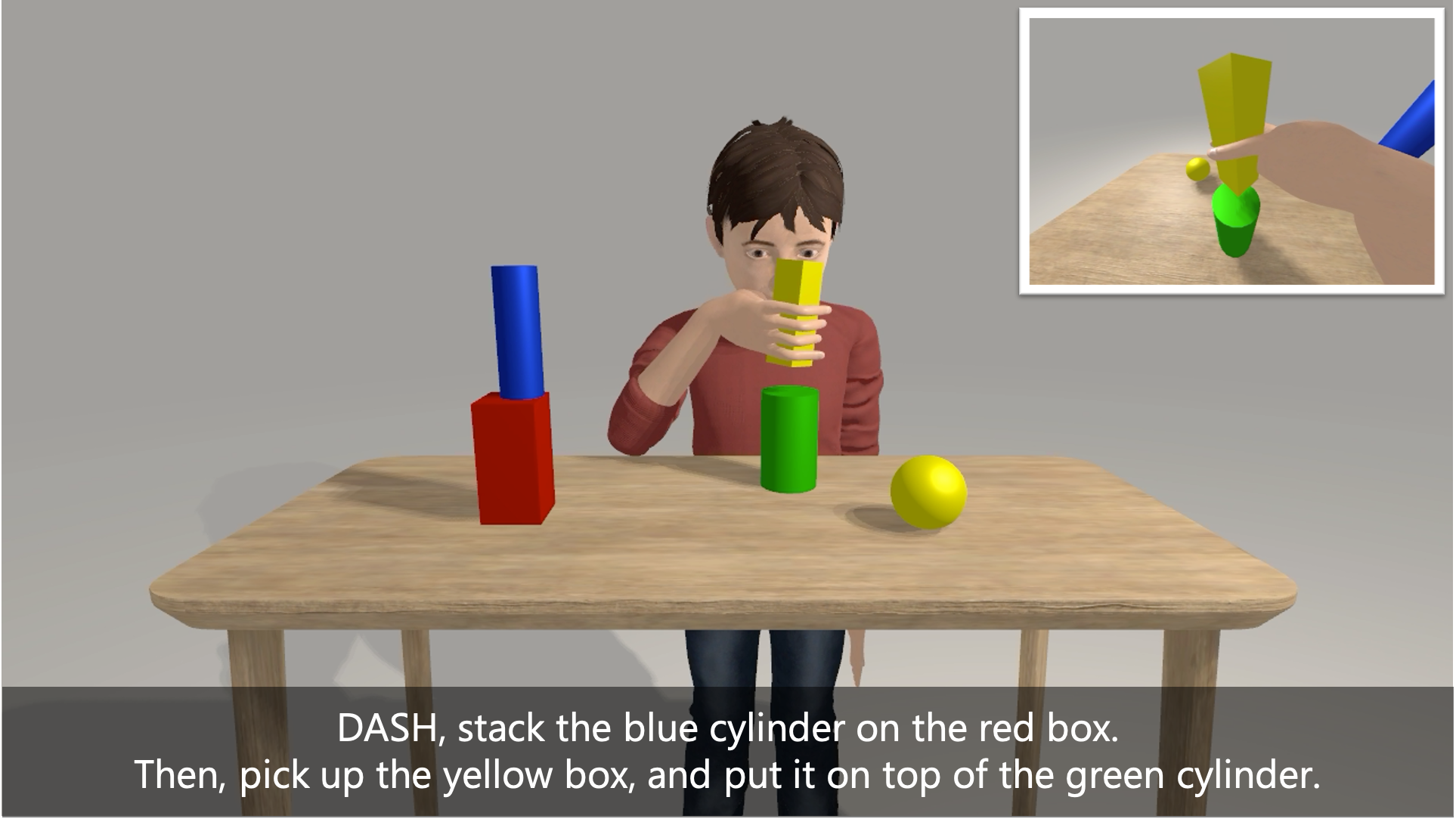}
\caption{Our system, \modelfull (\model), is an embodied virtual human modeled off of a child. \model is able to manipulate tabletop objects with a dexterous arm and hand in a human-like way, using its own vision perception (white box), language understanding (user command in black box), proprioception and touch. (Video 0m12s)}
\label{fig:pull}
\end{figure}


The recent breakthroughs in artificial intelligence (AI) have opened new avenues for computer graphics research \cite{render2020, 2018-TOG-SFV, chaudhuri2020learning}. Systems originally built for graphics applications are also playing an increasingly important role in the progress of AI.
For example, Embodied AI research, which centers around the idea that intelligence emerges from the inseparable interplay of sensorimotor activities and physical interactions with the environment \cite{smith2005development}, has utilized rendering systems and physics engines to provide inexpensive training data and testbeds. An integrated simulation environment with both realistic physics and visual appearances has the promise of creating diverse and safe playgrounds for virtual embodied AI agents to learn through interaction \cite{savva2019habitat, xia2020interactive, AllenAct}.



In this paper, we argue for the crucial inclusion of simulated humans in Embodied AI systems using integrated techniques and knowledge from Computer Animation. Such virtual humans should not only manifest realistic physics and appearance, but also function autonomously under embodied anthropomorphic constraints, such as egocentric vision, and with human-like interaction, such as natural language. Simulated humans that react to physics, perceive, and interact as real humans can provide an effective and safe platform for physical Human-Robot Interaction (pHRI) and assistive robotics, such as modeling human behavior during robot-assisted eating and drinking \cite{Erickson:2020}. Embodied virtual humans can also provide a new platform for scientific studies such as the perception-action coupling of infants' development: infants' initial physical interactions with the environment, though being imperfect with frequent failures, provide diverse experiences for perception development, which in turn facilitates motor skill learning. Though we envision most applications of our system to be in the virtual world, knowledge obtained through computational modeling of virtual humans could as well serve as a first step towards building physically embodied humanoid companions.



Such realistic modeling of human perception, actuation, and interaction constraints is in drastic contrast to existing virtual human systems, which either focus on visual appearance and motion realism \cite{10.1145/2933540.2933551, shapiro2011building}, or on higher-level vision and language capabilities while abstracting out actual motor skills \cite{thiebaux2008smartbody, hartholt2013all, puig2018virtualhome}. In comparison with current cognitive modeling systems, such as the ACT-R \cite{anderson1997act} simulator, which abstracts both perception and motor with symbolic modules, the physically embodied visual, verbal, and manual modules we aim to build could eventually be used to augment such systems. In contrast to existing character animation research, which usually does not simulate human perception, embodied virtual humans base their motor actions on noisy and partial perceptions of itself, rather than ground-truth simulator states.
 
To this end, we propose and build a system to physically simulate manipulation skills of a virtual human with egocentric vision, natural language command, proprioception, and touch---a first one to the best of our knowledge (Figure \ref{fig:pull}). We focus on object manipulation, which demands coordination between visual perception and motor dexterity. This system presents a unique set of requirements and challenges, as we summarize below:

\begin{itemize}
    \item \textbf{Generality}: A highly generalizable system is crucial for our envisioned applications, because a simple verbal command alone could generate a wide variety of scenarios---our system must be capable of manipulating objects of different shapes and sizes arbitrarily placed in the workspace, while avoiding an unknown number of surrounding, arbitrarily arranged objects. Our system should demonstrate high success rates on a large number of such scenes.
    
    \item \textbf{Anthropomorphism}: We extend the conventional notion of “being humanlike” in character animation research to include both anthropomorphic constraints and humanlike behaviors. Our system should operate under anthropomorphic constraints such as delayed and noisy visual input. Our system must demonstrate fluid motion on a high degrees-of-freedom (DoF) anthropomorphic arm and hand, in contrast to slow, conservative motion optimized for functionality that is typically seen on robotic manipulation systems. As we opt to explore many humanlike constraints to be imposed on our system, this work does not aim to create motions of better quality than current character animation techniques using full simulator states.
    
    \item \textbf{Extensibility}: Our system can be used as a self-contained black-box for Embodied AI research or pHRI. At the same, it should have sufficient modularity such that each module can be replaced or improved upon, so that more complex manipulation skills can be augmented to our first system without the need to redesign.
    
    \item \textbf{Interactivity}: The system should run at an interactive rate, allowing users to interact with the agent using online verbal commands.
\end{itemize}

Generality and anthropomorphism are often at odds with each other, because the demand of high success rates often results in a single conservative manipulation strategy with few natural variations. These important but sometimes conflicting goals prompt us to prioritize modularization in our system design. Specifically, we factor the problem into a vision module, a language module and manipulation modules for different stages, so that we can apply analytical methods to modules where correctness over a large range of scenarios are most important, and apply machine learning approaches to places where analytical methods are hard to solve and natural variations of motions are desired. As we are building a first system of its kind, modularization also facilitates analysis, interpretation, and extensibility.

We name our system Dynamic and Autonomous Simulated Human (DASH) and model DASH's visual appearance off a 5-year-old boy (Figure \ref{fig:pull}), as we believe modeling and understanding of the simultaneous acquisition of perceptual and motor skills are more important than demonstrating perfect skills themselves for embodied intelligence research. We evaluate the scenario in which DASH stands in front of a set of solid objects of various shapes and colors on a tabletop. The user gives DASH a sequence of grasp-and-stack commands in a domain-specific language to rearrange randomly scattered objects into a desired configuration. We evaluate the success rate of DASH accomplishing tasks correctly from randomized initial scenes, qualitatively demonstrate that DASH can produce fluid motions with natural variations, and analyze learned motor skills and various failure cases.

\section{Related Work}


\paragraph{Virtual Human Systems.} Creating virtual humans is a multidisciplinary research challenge in Computer Graphics, Artificial Intelligence, Cognitive Science, Human-Computer Interaction, and so forth. Many existing virtual human systems rely on pre-scripted, recorded or procedurally generated motion to animate the virtual humans in response to a set of predefined events \cite{Kalra:1998,Caicedo:2000,Magnenat-Thalmann:2006,Jung:2011,feng2014fast}. Multimodal dialog systems have demonstrated the use of virtual humans for verbal and nonverbal communication for training, education, and virtual assistant applications \cite{RJ:1999,Cassell:2000,Chi:2000,Lee:2006,nguyen2015modeling, matsuyama2016socially, perera2018building}. Computational modelings of human cognition \cite{anderson1997act, funge1999cognitive} and perception mechanisms \cite{peters2002synthetic, rabie2000active, Nakada:2018, eom2019model} have the promise of not only creating realistic motions but also answering why humans show certain behaviors. Our work similarly models realistic human perceptual and physical constraints, but further directly utilizing RGB images and physics simulations as input. Our system generates human motion from computational models rather than motion data, allowing generalizable, unprescribed interactions with the world and potentially answering scientific questions. 

\paragraph{Robotic Systems with Vision and Language.} Systems that can perceive visual scenes and understand language instructions in a humanlike way have been explored both in simulations and on real-world robots. Virtual agents in \cite{puig2018virtualhome, shridhar2020alfred} focus on high-level reasoning and abstracts motor skills with animated motion primitives, while our system focuses on fully physical motor skills. On real robots, similar manipulation systems \cite{paxton2019prospection, shridhar2018interactive, hatori2018interactively, liu2019review, jiang2018task, mees2020learning} usually employ robot grippers or suction cups where open-loop control primitives are robust enough for picking up objects. In contrast, this work explores the concatenation of multiple learned DRL controllers on a simulated anthropomorphic hand which produces more fluid and diverse motion while demonstrating robustness on precision tasks such as stacking of long and narrow objects.


\paragraph{Dexterous Manipulation in Computer Graphics.} Dexterous manipulation is an integral component to creating interactive virtual humans. In Computer Graphics, previous works have proposed to generate physically plausible human grasps from recorded motion capture data \cite{Pollard:2005,Zhao:2013}, videos \cite{Wang:2013}, or contact forces \cite{Kry:2006}, while our system aim at a more general grasp-and-stack task in a cluttered environment. Compared with other works \cite{Mordatch:2012, Liu:2009, Ye:2012, Andrews:2012} that animate dexterous manipulations, we rely on 2D images from the first-person view to infer estimated object information, without accessing contact point locations or forces from the ground-truth simulation states. A number of previous animation works also investigate the whole body motion during manipulation~\cite{Yamane:2004,Jain:2009,Ho:2010,Lee:2019, Ye:2012}. Merel \etal \cite{Merel:2019a, Merel:2020} recently developed a hierarchical control framework that utilizes egocentric vision for locomotion and whole-body manipulation. Our work focuses on complex dexterous manipulation tasks in cluttered environments which requires coordinated planning with vision and language inputs, without requiring motion data.

\paragraph{Dexterous Manipulation in Robotics.} In the Robotics community, manipulation primarily utilizes parallel jaw grippers, but dexterous hands have also been of great interest to researchers \cite{Bicchi:2000,Okamura:2000,Ma:2011}. While we also solve grasping with dexterous hands, our system lift the requirement of grasp contact planning, facilitating the learned manipulation policies to zero-shot transfer to novel geometries. In recent years, researchers have begun to investigate DRL approaches for dexterous manipulation \cite{Andrychowicz:2019,Akkaya:2019,Nagabandi:2019}. Our work is most closely related to \citet{Rajeswaran:2018,Radosavovic:2020}, where they used a combination of policy learning and imitation learning to solve a set of dexterous manipulation problems in simulation, including pick-and-place tasks. To reduce sample complexity and improve the quality of motion, Rajeswaran \etal \shortcite{Rajeswaran:2018} used a VR system and CyberGlove system to capture high fidelity demonstrations from real humans. In contrast, our work does not require high-quality human demonstrations acquired by specialized equipment. In addition, our pick-and-place must be cognizant of other objects in the scene so the hand can avoid collisions or interact with them.

\paragraph{Computer Vision and Natural Language Processing.} Recent development in deep learning has enabled perception systems that can directly learn scene representations from raw visual input, especially in constrained environments \cite{Kulkarni:2015,Wu:2017,Nguyen:2019,Sitzmann:2019, mees2020learning}. Directly related to our work, \citet{Yi:2018} integrated an explicit visual representation with a language understanding module, leads to high accuracy on visual reasoning.

Semantic parsing algorithms learn to map sentences to logical forms via a knowledge base or a program~\cite{Berant:2013,Liang:2013,Guu:2017}. Recently, \citet{Rothe:2017} showed that formal programs can be effectively used to model human questions. For \model, we choose to integrate an off-the-shelf learning-based semantic parser \cite{Honnibal:2017} in our language module.

\section{DASH: \modelfull}

\begin{figure}[h]
\centering
\includegraphics[width=\linewidth]{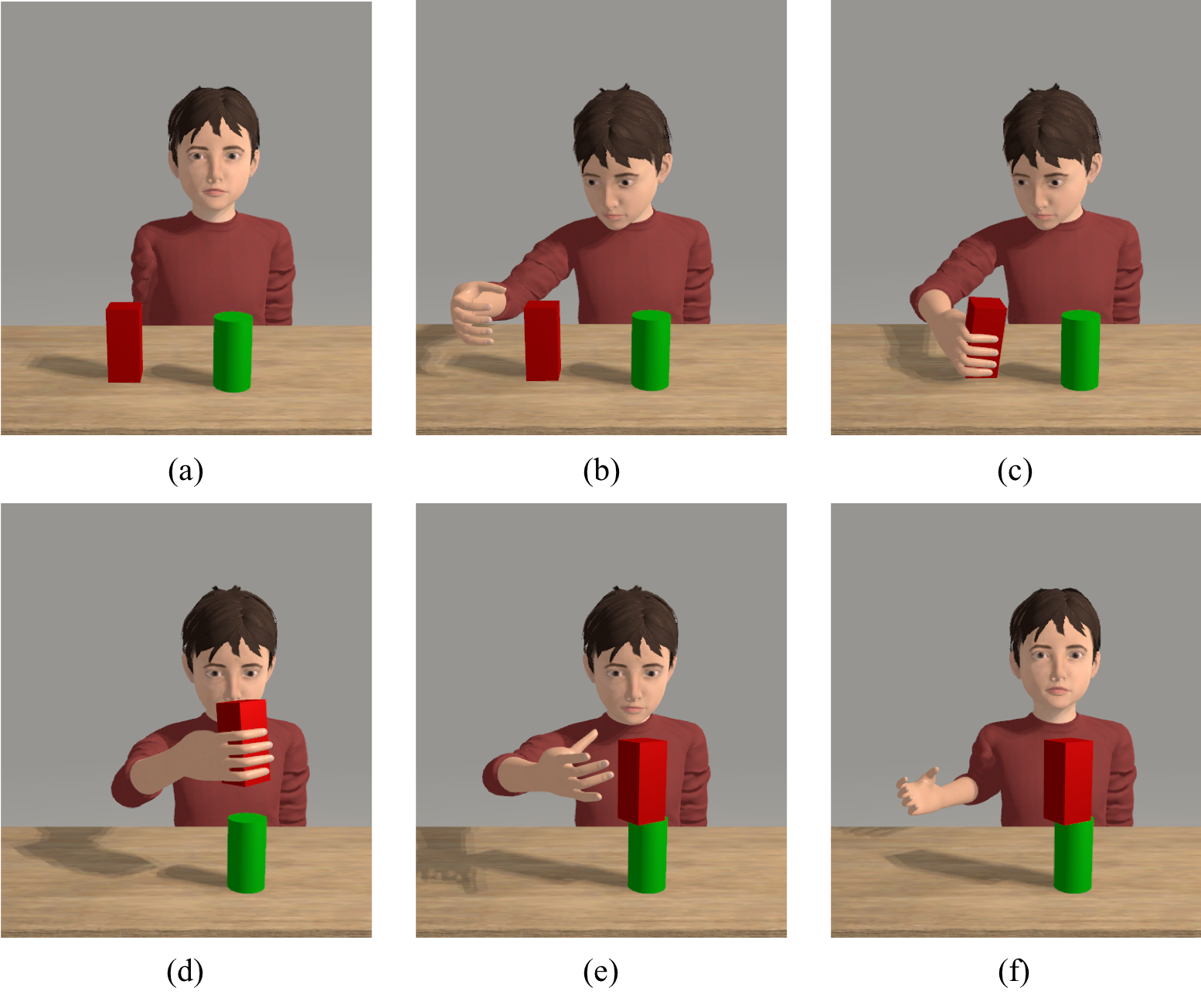}
\caption{We break down the general pick-and-place task into stages. (a) initial rest pose; (b) end of reaching stage; (c) end of grasping stage; (d) end of transporting stage; (e) end of placing/stacking stage; (f) retract to task completion pose.}
\label{fig:stages}
\end{figure}

We mix and match analytical and learning approaches to balance between generality and anthropomorphism requirements, and to favor extensibility. As such, the DASH system consists of an ego-centric vision module, a language parsing module and manipulation modules of two skill categories: transition and interaction. We decompose the manipulation tasks in this work into four self-contained stages: reaching, grasping, transporting, and placing/stacking (Figure~\ref{fig:stages}). We categorize reaching and transporting as transition stages, and grasping and placing/stacking as interaction stages. Here we discuss some key design decisions of our system.

\paragraph{Combined or independent vision and manipulation modules?} We design a standalone vision module so we can reuse the same training procedure for different purposes (e.g. initial planning, stacking). Such modular design also allows us to activate or deactivate vision module independent of the manipulation module. Compared with having one end-to-end neural module, separating vision from motor training also significantly reduces time necessary for training anthropomorphic arms and hands with high degrees of freedom (DoFs). We also construct the vision architecture to be object-centric, meaning that the same trained model should be able to estimate properties of a varying number of objects in scene.
 
\paragraph{Integrated or skill-specific manipulation modules?} Another design decision we made is to mix and match model-based planning and model-free reinforcement learning techniques for different manipulation stages. For transition stages (reaching, transporting), model-based planning has a probabilistic guarantee of a feasible solution if there exists one, agnostic to scene complexity; in contrast, learning a motor policy would require training to include all randomly arranged scenes, leading to a combinatorial explosion of sample demand. On the other hand, using learned controllers for interaction stages produces more fluid and diverse motions compared to conservative open-loop grasping and placing primitives usually seen for robotic grippers, which often manifest one single strategy with sole finger motion and the arm fixed throughout. Concatenating multiple policies in one manipulation system strikes a balance between functionality and anthropomorphism, but it also poses additional challenges much unexplored in previous works, which we discuss in Section \ref{sec:method-interaction}.
 
\paragraph{With or without human motion data?} A final key design choice is avoiding the dependency on motion data. For our envisioned applications where scientific questions, such as how impacted perception affects motor skills, are asked, a system mimicking motion data from healthy actors will not be sufficient. Besides, since we are interested in a wide variety of highly constrained motion, such as stacking thin objects in a cluttered environment, imitation learning alone might not be able to produce feasible motions. On the other hand, we found that natural reaching motion can be achieved using model-based path planning with re-timing based on humanlike velocity profiles.



\subsection{Vision Module}

\model includes a vision module that serves two main purposes: (i) to generate the scene observation for initial task planning and (ii) to provide frequent feedback to the manipulation module during object placing/stacking, both from raw RGB images. Our vision module consists of an object segmentation model and a feature extraction model, similar to \cite{Yi:2018}. Given a single scene image, the vision module first predicts segments for all objects in the scene. Each object segment is then concatenated with the original scene image and fed into a feature extraction network to output attributes of this object, making the model agnostic to number of objects in the scene. The model predicts discrete attributes such as shape and color, as well as continuous attributes including height, 3D position, and upright axis for each object (full orientation is ambiguous for symmetrical objects). Gaze (head movement) is animated to match the decided camera locations during stages (i) and (ii), and interpolated between them for stages when vision module is turned off. Replacing head movement with eye movement where possible could benefit overall naturalness.

While our manipulation module can generalize to novel geometries unseen during training (Video 3m14s), zero-shot transfer is much more challenging for learning based vision algorithms. We therefore leave the generalization of vision beyond primitive shapes for future work. The generation of training data will be discussed in detail in Appendix \ref{sec:vision-language-details}.

\subsection{Language Module} 
\label{sec:method-language}
The language module is invoked during initial task planning to infer the task parameters from the natural language command. We do not assume the specific form of the language command to allow natural variation in wording and ordering, as long as the command semantically specifies a target object to grasp, with reference object(s) and the desired final spatial relation to them. We assume both the target and reference objects can be modified by a shape, a color, and a modifier clause containing one secondary reference object and one secondary spatial relation for disambiguation.

Our language module consists of two steps: dependency parsing and parse tree search. We first use a robust neural-net-based dependency parser \cite{Honnibal:2017} to obtain the semantic dependencies within our instruction sentence in the form of a tree. This step is general to any sentence structure. For the second step, with our assumptions, we perform a breadth-first search on the parse tree to identify the target object to grasp, the desired spatial relation defining the movement, and reference object(s).

The outputs of the vision and language modules are then cross-referenced. There might exist multiple objects sharing the same shape and color in the visual scene. A secondary search is performed in the clause subtree that modifies the target or reference object. The system will halt if the instruction still remains ambiguous after this disambiguation step. Coordinates of the target object location and the placing destination location are sent to the dexterous manipulation module.

\subsection{Manipulation Module: Transition Stages}

The transition stages (reaching and transporting) can be solved with model-based motion planning, using a common procedure agnostic to scene arrangement:

\begin{enumerate}
    \item Compute the final arm pose that moves the hand close to target position for next stage, using the task parameters extracted by the vision and language modules;
    \item Plan a collision-free trajectory for the arm and hand (with the object in-hand in case of transporting);
    \item Execute the planned trajectory with close-loop tracking control.
\end{enumerate}

For our system, instead of finding a single destination pose, we produce a distribution of destination poses, all of which have the hand being close to the target position but with different palm facing orientations. Having multiple choices for the final pose increases the chance for the motion planner to yield a collision-free trajectory, alleviating failure cases where the destination arm configuration is already in collision. Moreover, the resultant motion exhibits richer diversity as the hand has multiple ways to approach the object. The distribution of poses are solved with a parameterized inverse kinematics (see Appendix \ref{sec:implement-transition}).

We solve for the collision-free trajectory with BiRRT algorithm \cite{Kuffner:2000}, followed by shortening, smoothing \cite{hauser2010fast}, and re-timing (Appendix \ref{sec:implement-transition}) to eliminate robot-like disjunct motions. The re-timed trajectory is physically tracked with a PD control law and inverse dynamics.

\subsection{Manipulation Module: Interaction Stages}
\label{sec:method-interaction}

We want to develop generic grasping and placing/stacking controllers capable of picking up objects of varying sizes, shapes, and locations, and placing objects on other objects or surfaces. We apply model-free deep reinforcement learning (DRL) to handle non-differentiable contacts involved, and fuse multiple sensory modalities of vision, proprioception, and touch. Our manipulation module can work without reasoning about grasping points, allowing it to zero-shot generalize to irregular object shapes unseen during training (Video 3m14s). We train one single grasping and one single placing/stacking policy by randomizing target object shapes and locations during training. Such randomization naturally leads the same policy to produce diverse behaviors. 

Crucial to a successful concatenation of different stages, the initial distribution of arm and finger pose during policy training should adequately cover the space of possible scenarios that may result from its preceding stage. For both grasping and stacking stages, the initial arm poses are generated using the same procedure for generating final arm pose distributions during reaching and transporting. However, there is an additional difficulty in designing initial distribution for the stacking policy due to a unique challenge of multi-finger hand manipulation. With a dexterous hand, the relative transform between the hand and the object-in-hand has a considerable variance among different grasps. Thus, to learn a robust placing/stacking policy, we need to include such variance in the initial state distribution. As such, we use the grasping policy to collect $10{,}000$ stable grasps, which we use to randomly initialize the hand pose and the configuration of the object-in-hand at the beginning of each training episode for the placing/stacking policy. Small white noise is injected to capture likely object slips during transportation.

Vision perception is crucial to placing/stacking, because this task requires updated feedback about the object configuration in the context of the scene. On the other hand, vision perception is less important for grasping as the reaching stage already leads the hand close to the object using the vision-estimated position from the initial scene. Thus, we use vision perception for placing/stacking and depend solely on proprioception and tactile perception during the grasping stage. Specifically, the observation space for the placing/stacking policy additionally includes the current 3D position and orientation of the involved objects as estimated by vision. For orientation, we only require the direction of the axis that indicates the upright position of the object (for spheres we simply use a unit-z vector). To mimic the latency and low frequency of human visual perception system, we update the vision input at $20$Hz and add $50$ milliseconds of delay in the vision input to the policy. Since the training data of vision module is generated after the stacking policy is trained, during training of stacking we simply feed the policy with true object poses from simulation.

Technical details of the DRL modules are discussed in Appendix \ref{sec:interaction-tasks}.

\section{Experiments}\label{sec:experiments}

In this section, we first evaluate whether the \model system can balance between robustness and anthropomorphism, for which we both `stress test' our system on $200$ randomly generated trials and evaluate the success rates of the full system, and report qualitative evaluation on interactions with \model. We also report the performance statistics of \model.

We then conduct experiments on \model's vision module and manipulation module to evaluate our key system design decisions. We provide analysis on the accuracy of vision module, showing that a standalone vision module facilitates analysis and interpretation. We compare with the alternative design of a manipulation module, showing that merging all stages in one policy would cause difficulty in learning. We additionally show the zero-shot transfer capability of our manipulation module to objects with unseen, irregular shapes, showing the advantage of using close-loop, partially observable DRL policies which require no contact point reasoning. We finish this section with qualitative examples showing our manipulation module produces smooth motions with natural variation, and can be easily extended to manipulation with the left arm and hand. 


\subsection{Evaluations of the Whole System} 

We compare the system performance between using (i) ground truth object states (upper bound) and (ii) vision-predicted object attributes as the input observation to \model using 200 randomly generated scenes (see Appendix \ref{sec:quan-setup} for details of setup). For ground truth evaluation, \model is provided with perfect object attributes and 3D pose from the PyBullet simulator. For vision evaluation, \model is provided with imperfect estimations of the objects.

\paragraph{Quantitative results.} We show (Table \ref{tab:success}) that \model achieves high success rates on the large number of random scenes. We report two metrics of success: (i) success when considering all trials, and (ii) success when excluding trials that encountered a motion planning failure.
A trial is considered successful if the $(x, y)$ location of the placed object is within 10 cm of the target $(x, y)$, and the $z$ coordinate of the center of mass is at least 5 cm above the target $z$ coordinate.
Success rates of \model when using vision-estimated observations are within 5.8\% of the success rates that use ground truth object states.
Stacking success rates are generally around 10 to 15\% lower than placing success rates, potentially due to higher sensitivity to observation accuracy.

\begin{figure}[b]
\centering
\includegraphics[width=0.85\linewidth]{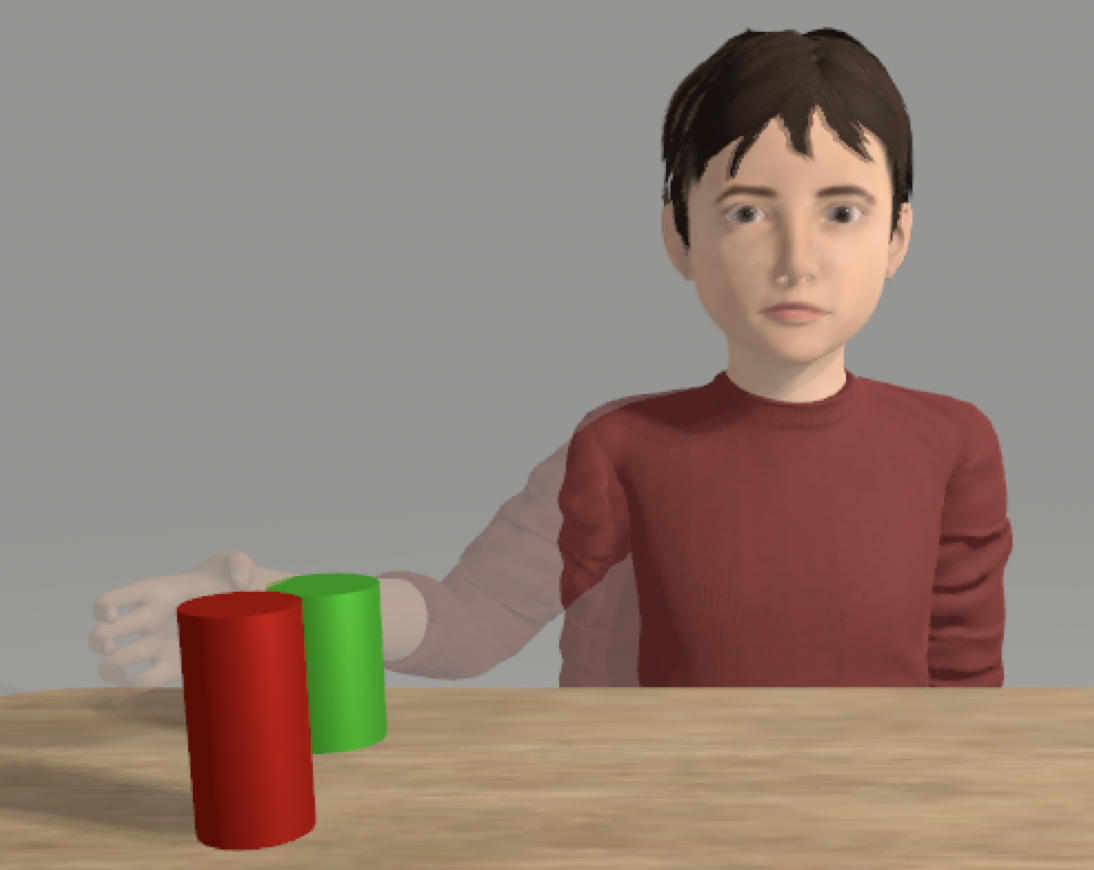}
\caption{Motion planner during reaching stage unable to find a collision free path to destination arm pose shown in transparent.}
\label{fig:motion-plan-error}
\end{figure}

We provide success rates excluding motion planning failures as additional reference since our randomly generated scenes and tasks may sometimes lead to motion planner unable to find a collision-free path during transition stages (Figure \ref{fig:motion-plan-error}). Potential improvements could be training \model to grasp from top of objects, and adding more Degrees of Freedoms (DoFs) around the shoulder complex. In summary, motion plan failures and stacking policy failures are the main sources of errors of our framework.

\begin{table}[t]
\small
\caption{\label{tab:success} Success rates (\%) of the full system on 100 placing and 100 stacking tasks. We report success rates when (i) including all trials and (ii) excluding trials that fail due to motion planning (Figure \ref{fig:motion-plan-error}). Vision denotes using imperfect estimations of object attributes from the vision module as input to DASH. Ground truth denotes perfect observation of object attributes. (Video 1m10s-1m35s)}
\begin{tabular}{lcccccc}
\toprule
                     & \multicolumn{3}{c}{All Trials}    & \multicolumn{3}{c}{Excluding Plan Failures} \\ \cmidrule(lr){2-4} \cmidrule(lr){5-7}
Observation          & Overall          & Place      & Stack      & Overall              & Place         & Stack         \\ \midrule
Vision               & 76.0             & 82.0       & 70.0       & 85.8                 & 89.1          & 82.4          \\
Ground truth         & 79.0             & 83.0       & 75.0       & 89.7                 & 91.2          & 88.2          \\ \bottomrule
\end{tabular}

\end{table}

\paragraph{Qualitative results.} To evaluate \model's motion quality, the accompanying video shows four successful examples of user-provided commands, mixing two-sentences and one-sentence commands with different kinds of modifiers (0m12s-1m10s). We wrote the instructions such that they cover the space specified by our assumptions (target object, reference object(s), at most one modifier clause), with variations in natural languages such as ``stack on'' vs ``put on top of''.
Example failure cases are also provided in the accompanying video (2m35s-3m02s).

\paragraph{Performance.} Performance of the DASH system is measured on a personal desktop with Intel Core i7-9700K processor and The NVIDIA GeForce RTX 2080 Ti graphics card. The whole system runs at around 58 simulation steps per second. Rendering and communications between PyBullet and Unity contribute to a significant portion of the running time. Without Unity rendering, the manipulation module runs at around 360 simulation steps per second. 


\subsection{Analyses of the Vision Module}

Having a stand-alone vision module allows us to analyze its performance independently. On the evaluation set ($20\%$ random split from the training data), as shown in Table \ref{tab:vision}, the vision module errors during initial task planning are much lower compared to the placing/stacking stage.
This is likely due to the fact that \model uses a fixed camera pose during planning, but varying camera pose to mimic human looking at destination during placing/stacking.
In addition, \model's arm is not in the camera view during the planning stage.
During the placing stage, \model's hand is in view, partially occluding the object in hand as well as surrounding objects, making pose estimation more challenging.
Finally, during placing, the vision module and the policy interact in a tight feedback loop, resulting in potential compounding of errors over time.
We also note that errors are generally lower when considering only successful trials instead of all trials.
This suggests that the vision module could occasionally encounter states from failed placing attempts that are outside of its training distribution.

\begin{table}[t]
\small
\caption{Vision module performance on the full system (test). We report classification accuracy on shape and color attributes, as well as the mean average error (MAE) on 3D position estimation (cm) across all predicted frames for the purpose of initial task planning, placing, or stacking. We show position prediction errors on all trials as well as when considering only successful trials.}
\begin{tabular}{lcccc}
\toprule
      & \multicolumn{2}{c}{Accuracy (\%)} & \multicolumn{2}{c}{3D Position MAE (cm)} \\ \cmidrule(lr){2-3} \cmidrule(lr){4-5}
Stage & Shape                & Color               & All Trials            & Successful Trials      \\ \midrule
Task Plan  & 100.0                  & 100.0                 & (0.8, 1.1, 0.4)       & (0.8, 1.3, 0.4)        \\
Place & 98.4                & 99.8               & (3.8, 3.2, 2.3)       & (2.8, 2.7, 1.8)        \\
Stack & 95.1                & 98.3               & (3.5, 2.9, 2.6)       & (2.7, 2.4, 1.9)        \\ \bottomrule
\end{tabular}
\label{tab:vision}
\end{table}

\subsection{Analyses of the Manipulation Module}

\paragraph{Comparison with learning-only approach.} We show that stacking all four manipulation stages in one policy would cause problems in learning (Video 4m01s). The learning-only approach formulates the entire pick-and-place problem as a POMDP and solves for a single policy to achieve all four stages of the task: reaching, grasping, transporting and placing. In our attempt to train such a policy on a dexterous hand without any motion demonstration, we found that the state-of-the-art learning method fails at the very beginning of the task. It was unable to discover a collision-free path to move the arm from a resting (straight down) position to one that is above the tabletop. Subsequently, we significantly reduced the difficulty of the problem to a manipulation task commonly addressed by prior art. The simplified task only includes reaching and grasping, starting from a simple initial state where a clear collision-free path is available. With such simplification, we were able to successfully learn a policy using a learning-only approach, similar to the findings from previous work. But on a similar simplified task which includes reaching, grasping and stacking in place the training still fails. These experiments confirm our hypothesis that a generic pick-and-place task ``in the wild'' can pose great challenges to a DRL approach, emphasizing the necessity of employing a hybrid approach of model-based planning and model-free learning.

\paragraph{Generalization to novel, non-primitive shapes.} 
Without any re-training, we further demonstrate the grasp-and-stack task on non-primitive objects including an hourglass, a cone, and a Stanford bunny (Video 3m14s-3m40s).
In these tasks we assume access to the object pose information without use of the vision module. Results show that our policy generalizes to novel, irregularly shaped geometries for the pick-and-place task.

\paragraph{Emergent behaviors with natural variations.} Since both our grasping and placing policies are required to handle a large range of task variation during training, multiple behaviors naturally emerge from the same grasping or placing policy (Video 1m40s-2m16s).
For instance, the grasping policy occasionally rotates the target object slightly in order to execute a more stable grasp.
In other cases, if an object is unstable and close to falling over during the grasping processes, the policy may switch to grasping the bottom of the object as a rescue strategy.
Similar variation of behaviors can be observed for the placing/stacking policy as well.
During placing or stacking, if the object in hand is in an upright orientation, the policy may choose to simply release the object from midair and allow the object to drop onto the placing surface.
In other scenarios, the policy may choose to be conservative and wait for the object placing to finish before releasing its hand.
If a placed object appears to be on the verge of falling over, the hand often maintains contact with object in an effort to restabilize it.

\begin{figure}[h]
\centering
\includegraphics[width=\linewidth]{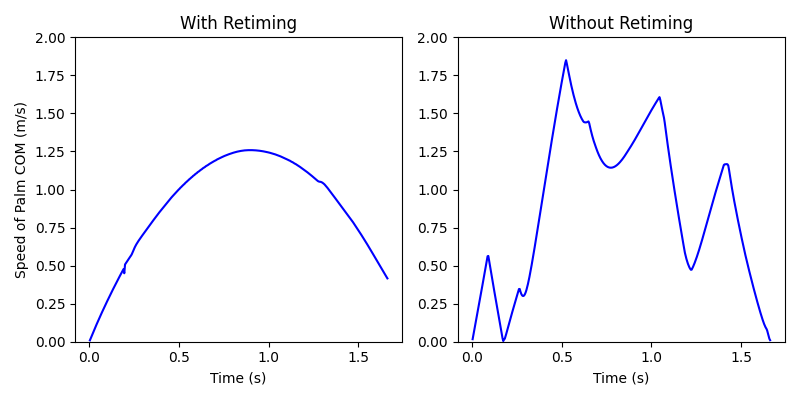}
\caption{Velocity profiles of the palm during reaching to object, with our retiming/reshaping technique, or using the OpenRave solved trajectory without editing.}
\label{fig:retiming-result}
\end{figure}

\paragraph{Effect of reshaping and retiming.} We show in the accompanying video (2m22s) that our reshaping and retiming technique (Figure \ref{fig:retiming}) creates more smooth reaching motions, compared with directly using the trajectory solved by the robotics package OpenRave. A comparison of the palm velocity profiles is also given in Figure \ref{fig:retiming-result}.

\paragraph{Extension: dual-arm manipulation.} In the accompanying video (3m45s), we show that the same policies trained for the right arm can be directly mirrored and used for the left arm.





\section{Discussion and Conclusion}

This paper introduces \model, a first system for simulating human manipulation skills with humanlike embodied constraints and language interaction. We show that it is possible to build an embodied virtual human that solely uses its own visual perception, natural language parsing, proprioception, and touch to perform grasp-and-stack tasks in a physically-simulated, cluttered environment. By factoring the \model system into a vision module, a language module and manipulation modules of two skill categories, we are able to mix and match analytical and machine learning techniques for different modules, so that \model is able to not only perform randomly arranged tasks with a high success rate, but also does so under incomplete information and with fluid and diverse motions. 

Our system does not depend on recorded human motion, hence having the potential to be used for both engineering applications such as physical human-robot interaction, and scientific applications such as the cognitive modeling of how embodied constraints affect human motions. By carefully concatenating DRL manipulation policies, our system can work without reasoning about grasping points, allowing it to zero-shot generalize to irregular object shapes unseen during training.

As the focus of the \model system is to explore the inclusion of various humanlike constraints, we are not able to use certain existing techniques to improve motion quality, such as grasping algorithms that utilize explicit geometry and contact point analysis. That said, the motion quality of the current version of \model can be much improved by allowing the torso to move and adding more DoFs at the shoulder complex. This will also enlarge the reachable area \cite{rodriguez2003bringing}, and could help the inverse kinematics solver to avoid finding poses with large wrist bending.

There are many possible choices for the vision module design. We choose egocentric RGB vision as it is closer to the human vision system. Alternative designs such as third-person point clouds reconstructed from dual cameras could have made the system easier to build. More scientific knowledge can be injected to the vision module to make it more humanlike. For example, the location uncertainty of human vision is not only distance dependent, but also depends on nearby reference objects including the reaching hand.


\model is currently limited to grasping and placing/stacking tasks, but the modular design facilitates extensions such as concatenating additional DRL policies for more complex skills. Just as humans simultaneously learn perception and motor skills, interleaving stacking and vision training (instead of only once in the current system) could further improve the robustness of the system. 

Even though the use of partially-observable DRL policies allows our manipulation module to transfer easily to unseen shapes, achieving the same transfer for vision module is much more challenging. One future direction is to train the vision module with a large set of everyday objects, randomly arranged on the tabletop. Finally, while our language module is robust to some wording (\eg ``put on'' versus ``stack it on'') and natural ordering variations, it will not handle ambiguous commands. This can be addressed by allowing DASH to ask domain-specific questions when instructions are ambiguous.




\begin{acks}
The authors would like to thank Michael Hayashi for mirroring the trained policies for the dual-arm setting.
\end{acks}
\bibliographystyle{ACM-Reference-Format}
\bibliography{references}

\newpage

\appendix

\section{System Implementations}
In this section we present additional technical material that is important to the success of the DASH system.


\subsection{Vision and Language Modules}
\label{sec:vision-language-details}
\subsubsection{Vision module.} We use Mask R-CNN~\cite{he2017mask} with a ResNet-50 FPN~\cite{lin2017feature} backbone as our segmentation model, implemented using the ``Detectron'' library~\cite{girshick2018detectron}.
The model is trained for 100K iterations with a batch size of two and learning rate of $2.5 \times 10^{-4}$.
For our feature extraction model, we train a ResNet-18~\cite{he2016deep} for each of the planning, placing, and stacking stages for 60K iterations using a learning rate of $0.002$ and batch size of 20.
The model inputs are resized to 480 by 480.

The model is trained in camera coordinates; at test time, the predictions from the vision module are transformed from the camera coordinate frame into the world coordinate frame using the camera pose. We then use the Kuhn-Munkres (Hungarian) algorithm~\cite{kuhn1955hungarian} to associate objects predicted across time, based on their predicted attributes.

We use the Unity 3D game engine\footnote{\label{footnote:unity}Unity game engine website: \url{https://unity3d.com/}} to render training images and object segmentations.  The camera is placed near the eyes of \model to provide vision observation. 
At the initial task planning (before manipulation starts), the \model camera points at a fixed location such that the camera view covers the main region that manipulable objects lie. The information predicted from this first image is used to generate a task plan jointly with language module.
During placing/stacking stage, the camera points at the destination $x,y$ location predicted during task planning. The camera target locations during training are computed using ground truth object locations with 2 centimeter noise for robustness.

The vision module is trained on frames from both the initial tabletop scene and the placing stage to serve its two usages.
We generate 20K training images of initial tabletop scenes following the same procedure as that of our full system evaluation pipeline (see Appendix~\ref{sec:quan-setup}).
To generate examples from the placing stage, we collect 4,000 rollouts from a pretrained placing/stacking policy.
For this step of data generation, the policy utilizes perfect object pose information from simulator, with an additional 2 centimeter noise on object positions and 0.03 noise on up vectors. From all frames generated from the policy rollouts, we uniformly downsample 40K frames for vision training.

\subsubsection{Language module.} We used the neural network based dependency parser proposed by \citet{kiperwasser2016simple} which is robust to noise in natural language, and its open source implementation in spaCy~\cite{Honnibal:2017}. For the breadth-first search, we pre-define a dictionary of shape, color, and spatial relation keywords. There might exist multiple objects sharing same shape and color in the visual scene. To disambiguate, we then search in the subtrees of the target object and of the primary reference object token for secondary spatial relations and secondary reference object(s) that modify the target or reference object.

\subsection{Manipulation Module: Transition Stages}
\label{sec:implement-transition}
The manipulation controller operates on an anthropomorphic arm and hand including four fingers, a thumb, and a palm, totaling $29$ degrees of freedom. For clarity of exposition, we will refer the degrees of freedom (dofs) from the shoulder to the palm as $\vc{q}^{A} \in \mathbb{R}^7$ and the dofs for finger joints as $\vc{q}^{F} \in \mathbb{R}^{22}$. We only simulate and control \model's right arm and right hand in main experiments, but also show that the same manipulation controller can be easily mirrored to the left arm and hand. We use the PyBullet physics engine~\cite{coumans2019} as the simulator. 

\subsubsection{Compute the final arm pose.}
From the initial image of the scene and the language instruction, the vision and language modules jointly infer the high-level pick-and-place task parameters, including the index of the target object, its initial position $\bar{\vc{p}}_i \in \mathbb{R}^3$, its desired position $\bar{\vc{p}}_d \in \mathbb{R}^3$, and estimated initial locations and shapes of all objects in the scene. Our first goal is to find the final arm pose $\bar{\vc{q}}^A_T$ such that the palm is close to the initial position $\bar{\vc{p}}_i$ at the end of the reaching stage, and close to the desired position $\bar{\vc{p}}_d$ at the end of the transporting stage.
 
To solve a distribution of final poses for the arm, we consider an Inverse Kinematics (IK) problem that matches the position and orientation of the palm in the world frame to a desired affine transformation $\vc{T}$, parameterized by a random variable $\theta$,
 \begin{equation}
     \vc{T}(\theta) = \begin{bmatrix}\vc{R}_z(\theta) & \vc{p} \\ \vc{0} & 1\end{bmatrix} \vc{T}_0,
 \end{equation}
where $\vc{p}$ is assigned to the initial or desired position $\bar{\vc{p}}_i$ or $\bar{\vc{p}}_d$ depending on the stage, $\vc{R}_z(\theta)$ is a rotation matrix about z-axis (upright axis) by angle $\theta$, and $\vc{T}_0$ is a constant transformation from the object frame to the palm frame, manually designed for each task. For example, for the reaching stage, $\vc{T}_0$ positions the right palm at $20$cm to the right of the object, with the palm facing the object. We uniformly sample $\theta \in [0, 2\pi]$ to create multiple candidates for the final arm pose. The implication is that we can grasp and hold the object from multiple orientations by its side, but not from its top. This multiple candidate IK problem is iterated by candidate and each solved with \cite{buss2004introduction}. The selected final arm pose $\bar{\vc{q}}^A_T$ is a collision-free pose with the wrist closest to its rest angle.


\begin{figure}[h]
\centering
\includegraphics[width=\linewidth]{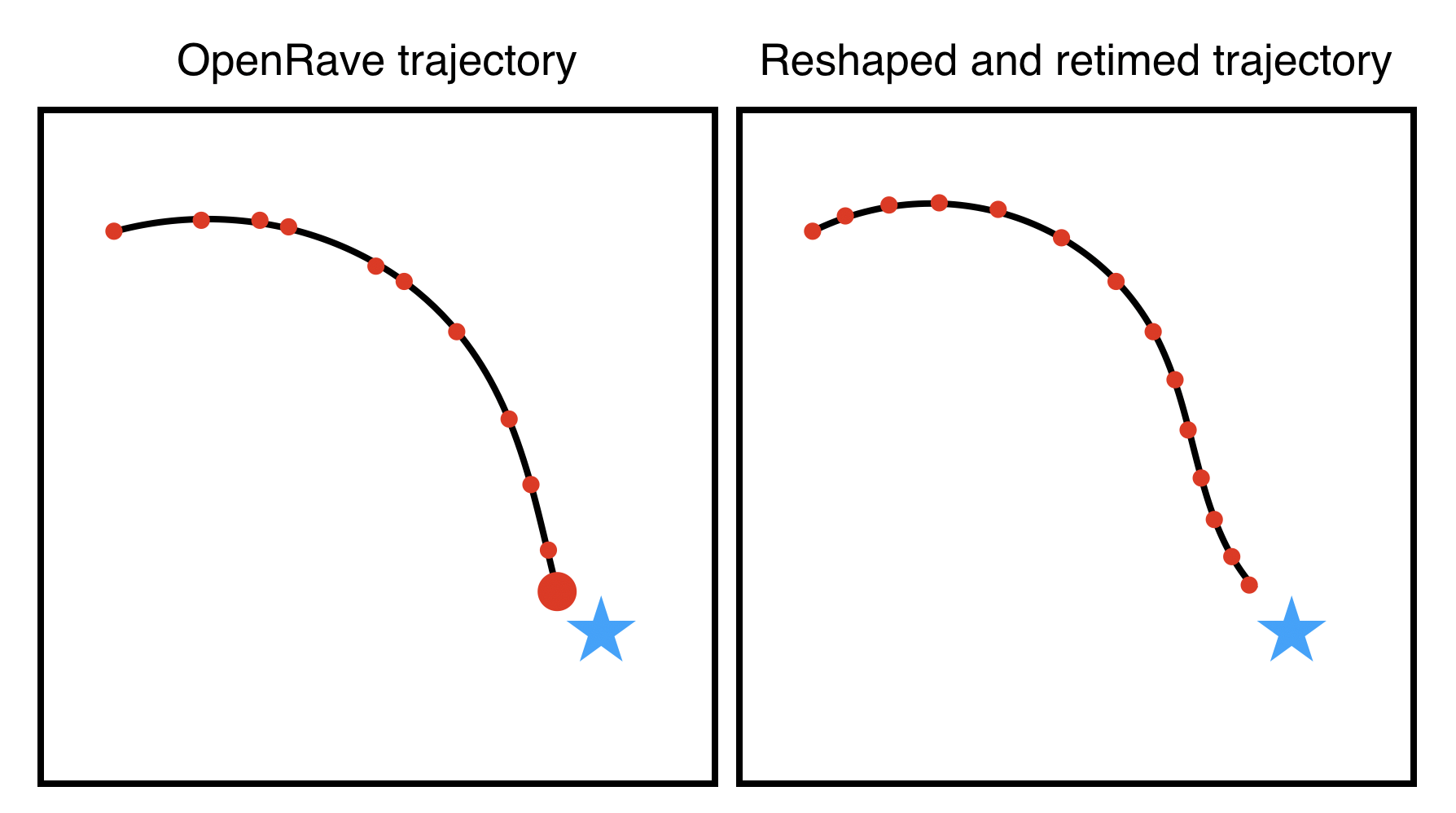}
\caption{Illustration: Reaching trajectory of the arm/palm solved from OpenRave is reshaped and retimed to have non-zero velocity before the grasping motion starts, making the stage switching more natural.}
\label{fig:retiming}
\end{figure}

\subsubsection{Plan a collision-free trajectory.} Once the final arm pose $\bar{\vc{q}}^A_T$ is determined, the next task is to plan a smooth trajectory for the arm and the hand from the initial pose given by the previous stage (for the first reaching stage we simply use the rest pose) to $\bar{\vc{q}}^A_T$, while avoiding collision with objects in the scene.

There exist many efficient and reliable motion planning algorithms from robotics literature. We apply BiRRT~\cite{Kuffner:2000} to first search for a collision-free path and then shorten and smoothen the path using techniques proposed by \cite{hauser2010fast}. We use the open-source implementation in OpenRave~\cite{diankov_thesis} to obtain a collision-free trajectory $\bar{\vc{q}}^A_{1:T}$.

The trajectory solved by OpenRave assumes zero terminal palm velocity since this is the usual convention in robotics systems. While this is fine for the transition between transporting and stacking, complete stop of palm at the switching from reaching to grasping would result in unnatural, robot-like motions. As OpenRave does not control the direction of the terminal tangent (because of zero terminal speed), we first reshape the OpenRave trajectory by replacing the final 10\% with a Hermite curve, so that the terminal tangent points from the palm towards the object; then a retiming is performed on the edited trajectory to fit a quadratic velocity profile, mimicking that human arm first accelerates then decelerates during reaching \cite{krishnan2017can}. We set the terminal palm speed to $0.4\textrm{m/s}$. Figure \ref{fig:retiming} shows an illustration of this procedure. 


The finger joints trajectory $\bar{\vc{q}}^F_{1:T}$ holds the same manually designed pose in Figure \ref{fig:stages} (b) throughout reaching, and the same grasping pose (see next paragraph) throughout transporting. During planning, we replace the hand (and the object-in-hand) with a conservative bounding box to speed up the computation.

\subsubsection{Simulate the planned trajectory.}
To track the planned trajectory $\bar{\vc{q}}^A_{1:T}$ in a physics simulation, we first compute the desired velocity $\dot{\vc{q}}^+$ in the next time step using a PD control law,
\begin{equation}
\label{position_control}
    \dot{\vc{q}}^+ = -K_p (\vc{q}_t - \bar{\vc{q}}_t) - (K_d - 1) \dot{\vc{q}}_t,
\end{equation}
where $\vc{q}_t$ and $\dot{\vc{q}}_t$ are the joint state at current time step. The desired velocity is realized by solving for the appropriate control torque via Inverse Dynamics, and then sending solved torques to joint motors for forward simulation in Bullet. Note that $\dot{\vc{q}}^+$ is not always achieved due to joint limits, torque limits, and the presence of environment contacts. The stability of simulation is not sensitive to the gains $K_p$ and damping coefficients $K_d$, as a result of solving torque from Inverse Dynamics.

During the transporting stage, we must ensure that the grasped object does not fall out of the hand.
Since the contact force applied to the object is due to the tracking error of the hand pose, the target hand pose should be designed in a way such that it induces effective forces to secure the object and is unachievable during the transporting phase (so the hand will continuously apply contact force to the object). Instead of manually designing such a pose, we use the sign of the tracking error in the last time step during the grasping stage, $\vc{e} = \sign(\vc{\bar{q}}^{F}_g - \vc{q}^{F}_g)$, to define the target pose during transporting as $\bar{\vc{q}}^F_{tran} = \vc{q}^{F}_g + 0.1 \vc{e}$.


\subsection{Manipulation Module: Interaction Stages}
\label{sec:interaction-tasks}

We train a policy $\pi(\vc{a}_t | \vc{o}_t)$ that maps the agent's current observation $\vc{o}_t$ to its action $\vc{a}_t$ every $k=6$ simulation steps. We formulate the problem of grasping and placing or stacking objects as a Partially-Observable Markov Decision Process    $(\mathcal{S}, \mathcal{O}, \mathcal{A}, f, r, p_0, g, \gamma)$, where $\mathcal{S}$ is the state space, $\mathcal{O}$ is the observation space, $\mathcal{A}$ is the action space, $f(\cdot)$ is the system dynamics, $r(\cdot)$ is the reward function, $g(\cdot)$ is the observation function, $p_0$ is the initial state distribution, and $\gamma$ is the discount factor. We use a policy gradient method, PPO~\cite{Schulman:2017}, to solve for a policy $\pi$ such that the accumulated reward is maximized:
\begin{equation}
    J(\pi) = \mathbb{E}_{\mathbf{s}_0, \mathbf{a}_0, \dots, \mathbf{s}_T} \sum_{t=0}^{T} \gamma^t r(\mathbf{s}_t, \mathbf{a}_t),
\end{equation}
where $\mathbf{s}_0 \sim p_0$, $\mathbf{a}_t \sim \pi(\mathbf{a}_t|\mathbf{o}_t)$, $\vc{o}_t \sim g(\vc{s}_t)$ and $\mathbf{s}_{t+1}=f(\mathbf{s}_t, \mathbf{a}_t)$.
Unlike model-based motion planning, reinforcement learning requires offline training of the policy. While the observations will be partial and noisy, we assume during training the reward function additionally has access to the complete, accurate simulator state $\vc{s}_t$ to better evaluate training progress.
 
\subsubsection{Observation space and action space.}
The observation space contains information from proprioception and tactile sensors, as well as information about the task. The stacking policy additionally uses frequent vision module estimates, as discussed in Section \ref{sec:method-interaction}.

The proprioception consists of the current state and the target pose from the previous time step: ($\vc{q}_t$, $\vc{\dot{q}}^{A}_t$, $\vc{\bar{q}}_{t-1}$). Empirically, we found that finger velocity $\vc{\dot{q}}^{F}_t$ does not improve performance, and its wide range and sensitivity to contacts hurts generalization. The tactile perception on the hand is represented as a vector of binary numbers $\vc{c}_t$, each of which denotes whether the corresponding phalanx is in contact with an object. The task information contains the location of interaction, \ie, $\bar{\vc{p}}_i$ for grasping and $\bar{\vc{p}}_d$ for placing/stacking, and a one-hot vector representing the shape of the object (e.g., cylinder, box, or sphere) to grasp, place, or stack.


The action space is defined as the change in desired joint angles $\Delta \bar{\vc{q}}$ from the previous time step. Using the same velocity-based control scheme described in Appendix \ref{sec:implement-transition}, we compute the required torque so that the finger joints can closely track $\bar{\vc{q}}_{t} = \bar{\vc{q}}_{t-1} + \Delta \bar{\vc{q}}$.

\subsubsection{Initial state distribution.} At the beginning of each rollout during training, we sample from the initial state distribution $p_0$ to reset the scene containing the arm, hand and involved objects, from which the agent starts a new trial of grasping or placing/stacking. Our goal is to design $p_0$ such that it covers the space of possible scenarios that may result from its preceding stage. As such, $p_0$ first provides randomization over the involved object's shape, size, mass, and friction coefficient, sampled from uniform distributions with ranges listed in Appendix \ref{sec:quan-setup}.
We also sample $\bar{\vc{p}}_i$ or $\bar{\vc{p}}_d$ uniformly to cover the reachable range of the arm.
We apply white noise within $2$ cm to the sampled $\bar{\vc{p}}$ and use it to reset the involved object's initial position. The additional white noise prepares the policy to handle vision inaccuracy of $\bar{\vc{p}}$ from initial task planning. With the updated $\bar{\vc{p}}_i$ or $\bar{\vc{p}}_d$, we follow the procedure described in \sect{sec:implement-transition} to generate a set of arm poses parameterized by $\theta$.



\subsubsection{Reward function design for grasping.}
Empirically, we found that the more variation in the initial state distribution that our policies are required to handle, the more carefully we need to design the reward function in order to help policy training avoid undesired local optima.
The reward function of our grasping policy at each step $r^G_t(\vc{s}_t, \vc{a}_t)$ consists of the following terms (all norm operators $\norm{\cdot}$ below denote Euclidean distance):
\begin{equation}
  r^G_t = 10r^o_t + r^p_t + 2r^c_t + 1.5r^s_t + r^d_t,
\end{equation}
each of which we will elaborate below.

The object pose term $r^o_t$ discourages objects from moving too much horizontally, which might lead to knocking over surrounding objects during test time: 
\begin{equation}
    r^o_t = -\norm{(\vc{p}_t - \bar{\vc{p}}_i)_{x,y}},
\end{equation}
where $\vc{p}_t$ is the object location provided by the simulator and the subscript $(x,y)$ indicates that the height difference is ignored.

Let $\vc{h}_1(\vc{q}^F), \cdots, \vc{h}_5(\vc{q}^F)$ be the world locations of the finger tips from the thumb to the little finger, and $\vc{h}_p(\vc{q}^F)$ be the world location of the palm. The form closure term $r^p_t$ encourages five finger tips and the palm to stay close to the object:
\begin{equation}
\label{eqn:form_closure}
    r^p_t = -5 \norm{\vc{h}_{1,t} - \vc{p}_t} - \sum_{i=2}^5 \norm{\vc{h}_{i,t} - \vc{p}_t} - 2  \norm{\vc{h}_{p,t} - \vc{p}_t}.
\end{equation}

The contact term $r^c_t = |\vc{c}|$ encourages more phalanxes to be in contact with the object. Similarly to Equation \ref{eqn:form_closure}, we weigh the thumb contact five times more than other fingers.

The finger shaping term $r^s_t$ encourages the four fingers to have a similar pose:
\begin{equation}
    r^s_t = - \norm{\vc{q}^{F2}-\vc{q}^{F3}} - \norm{\vc{q}^{F3}-\vc{q}^{F4}} - \norm{\vc{q}^{F4}-\vc{q}^{F5}} -  \norm{\vc{q}^{F5}-\vc{q}^{F2}}, 
\end{equation}
where $\vc{q}^{F2}, \cdots, \vc{q}^{F5}$ are the subsets of $\vc{q}^F$ that correspond to the four finger DOFs.

Finally, similar to the training scheme used in \citet{merzic2019leveraging}, we add a `drop test' around the end of each training episode to increase robustness of the policy. During the test, we remove the table and apply downward external forces on the object for around 0.4s. We add a penalty $r^d_t = -15$ for each control step during this test if the object is dropped. Designing a reward function that encourages lifting behaviors can be done by rewarding force closure or terminal height of the object, but we found that directly altering the physics environment results in more sample-efficient learning.

\subsubsection{Reward function design for placing and stacking.} 
Designing the reward function for the placing and stacking policy is more challenging as success states are sparse and the dynamics are much less stable. At each step the reward $r^P_t(\vc{s}_t, \vc{a}_t)$ consists of the following terms:
\begin{equation}
  r^P_t = 5r^o_t + 10r^a_t - 0.5r^c_t + 0.3r^s_t + r^b_t.
\end{equation}

With a slight abuse of notation, the object pose term $r^o_t$ for placing penalizes both position and orientation deviations from a desired object final pose $\bar{\vc{p}}_f$, where the object-in-hand is in an upright pose above the bottom object or at the correct location on the tabletop:
\begin{equation}
    r^o_t = - \norm{\vc{p}_t - \bar{\vc{p}_f}} + 3\vc{z}^T\vc{R}_t \vc{z},
\end{equation}
where $\vc{p}_t$ is the position of the object being placed,  $\vc{R}_t$ is its rotation matrix, and $\vc{z} = [0,0,1]$, the direction of gravity.

The action sparsity term $r^a_t$ encourages sparse hand force during placing. Since in our PD-like control scheme, the source of hand force is the tracking error $\vc{\bar{q}} - \vc{q}$, $r^a_t$ is defined as: 
\begin{equation}
    r^a_t = 1 / (\norm{\vc{\bar{q}}_t - \vc{q}_t} + 1).
\end{equation}

We use the same definition of contact term $r_c$ and the finger shaping term $r_s$ as in those defined for grasping. Instead of encouraging contact with the object ($2r_c$), we slightly discourage ($-0.5r_c$) it during placing/stacking.

Finally, a bonus term $r^b_t = 5$ is added for each step \emph{if} the object-in-hand is sufficiently close to $\bar{\vc{p}}_f$, and an additional bonus $r^b_t = 20$ is added \emph{if} the hand releases object when the object is close to $\bar{\vc{p}}_f$.

\subsubsection{Training details.}
We use the default learning parameters reported in the original PPO algorithm~\cite{Schulman:2017} to train both the grasping and placing/stacking policies. We train the grasping policy for 12M simulation steps and placing/stacking for 16M simulation steps. To train one single policy that could both stack an object and place it on tabletop, we randomly split 70\% of the training episodes for stacking trials and 30\% for placing. Since stacking spheres is unlikely to be successfully trained, we trained a separate sphere grasping policy and placing-on-table policy for qualitative demonstrations only. 

\section{Quantitative Evaluation Setup}
\label{sec:quan-setup}
We randomly generate $200$ scenes and tasks to systematically assess the generality of our system. For each scene, the task is randomly selected from either (i) place an object at a specified $(x, y, z)$ location, or (ii) stack an object on top of another object. Language is excluded from this experiment, as random commands generated by a program is effectively the same as randomly sampled the parameters of pick-and-stack tasks. We leave user studies with natural language commands provided by users to future work.


For each object in the scene, we randomly sample shape (box, cylinder, or sphere), 
color (red, yellow, green, or blue), 
width ([6, 10] cm), 
height ([13, 18] cm), 
base position ($-0.1 \leq x \leq 0.25$, $-0.1 \leq y \leq 0.5$, $z=0$),
mass ([$1.0$, $5.0$] kg), and
lateral (linear) contact friction coefficient ([0.8, 1.2]).
Box widths are scaled by $0.8$ to roughly match the length of its diagonal with the diameter of its cylindrical counterpart.
We sample sphere sizes according to the object height sampling range, downscaled by $0.75$. 
The minimum distance between object centroids was set to $25$ cm.
We randomly generate between 4 and 6 objects per scene, subject to the constraints listed above, and resample if generated objects fail to meet our criteria.
We limit the number of resampling trials to 50.
As a result, some scenes may have fewer than the desired object count.


For stacking, we set the minimum width of the bottom object in the stack to be slightly thicker ($9$ cm) to avoid overly difficult tests of stacking onto a narrower object than self.
We additionally enforce shape and color uniqueness between the task objects and surrounding objects to ensure that the task specification is unambiguous without providing language commands.

\end{document}